# Women in Italian astronomy

*Francesca Matteucci, Università di Trieste, Chair of INAF Scientific Council*

*Raffaele Gratton, INAF-Osservatorio Astronomico di Padova*

## Summary
*This document gives some quantitative facts about the role of women in Italian astronomy. More than 26% of Italian IAU members are women: this is the largest fraction among the world's leading countries in astronomy. Most of this high fraction is due to their presence in INAF, where women make up 32% of the research staff (289 out of 908) and 40% of the technical/administrative staff (173 out of 433); the percentage is slightly lower among permanent research staff (180 out of 599, about 30%). The presence of women is lower in the Universities (27 out of 161, about 17%, among staff). In spite of these (mildly) positive facts, we notice that similarly to other countries (e.g. USA and Germany) career prospects for Italian astronomers are clearly worse for women than for men. Within INAF, the fraction of women is about 35-40% among non-permanent position, 36% for Investigators, 17% for Associato/Primo Ricercatore, and only 13% among Ordinario/Dirigente di Ricerca. The situation is even worse at University (only 6% of Professore Ordinario are women). We found that similar trends are also present if researchers are ordered according to citation rather than position: for instance, women make up only 15% among the 100 most cited astronomers working in Italy, a percentage which is however twice that over all Europe . A similar fraction is found among first authors of most influential papers, which cannot be explained as a residual of a lower female presence in the past. We conclude that implicit sex discrimination factors probably dominate over explicit ones and are still strongly at work. Finally, we discuss the possible connection between the typical career pattern and these factors.*

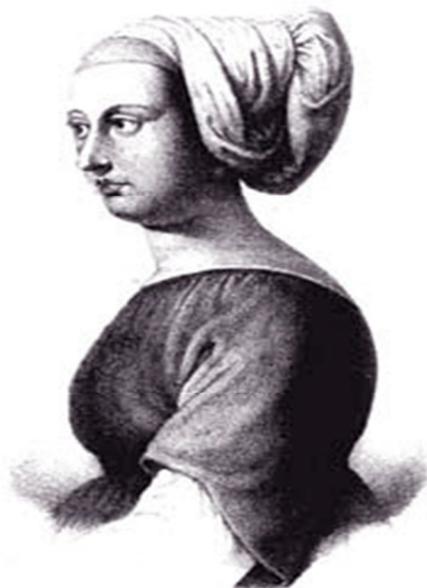 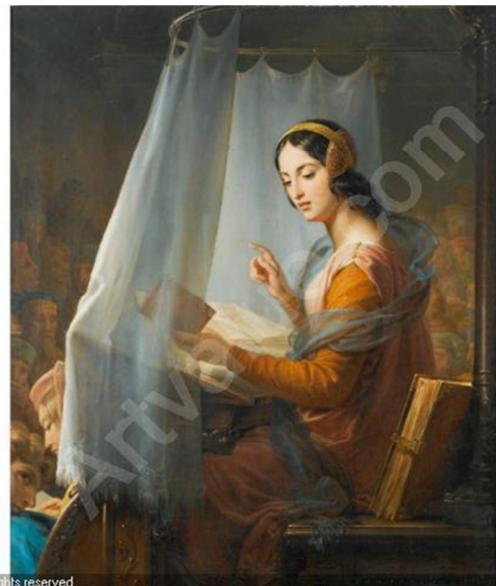

Figure 1. Left: Bettisia Gozzadini (1209-1261); right: Novella D'Andrea (1333-?)

# Introduction

The presence of women in science is hampered by several factors, both cultural and sociological. Only 2 Nobel Laureates in Physics out of 196 were women: Marie Sklodowska Curie (1903) and Maria Goeppert Mayer (1963)[1]. We are interested here to establish what the status of women is in Italian astronomy. Italy has a long tradition in science. Although rare, a few Italian women played a role in science since the middle ages: for instance, Bettisa Gozzadini (1209-1261) and Novella D'Andrea (1333-?) were the first women teachers at University of Bologna (see Figure 1). The first woman with a chair in astronomy was Margherita Hack (see Figure 2), who became professor at the University of Trieste and Director of the Trieste Observatory in 1964. Since then, the presence of women in Italian astronomy has very rapidly risen. In this document, we present some facts about women in Italian astronomy and briefly discuss their current status.

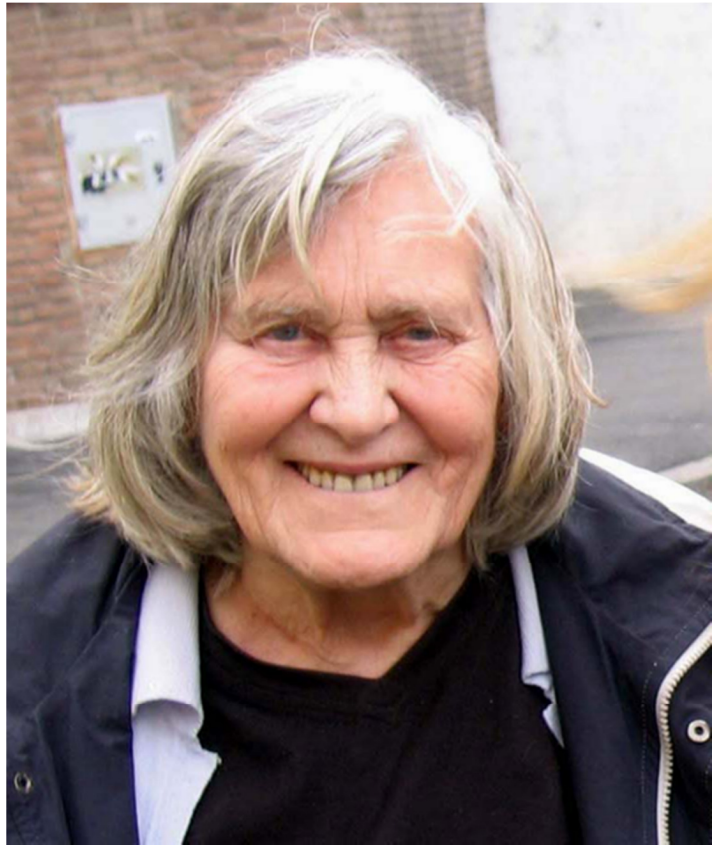

Figure 2. Margerita Hack (1922-2013)

# Women in Italian Academies

Only quite recently women have become common among scientific members of Italian Academies. The current situation is as follows:

Accademia dei Lincei: Section for Mathematical, Physical and Natural Sciences, 11 women on a total of 189 members (roughly 6%). A list of famous members of Lincei includes Albert Einstein, Guglielmo Marconi, Enrico Fermi, Rita Levi Montalcini and Margherita Hack.

---

[1] Source http://www.nobelprize.org/

Istituto Veneto di Scienze Lettere ed Arti (Venetian Institute for Sciences, Literatures and Arts): total 280 members, 24 women (8.6%).

In addition, we notice that there are 40 Italian women among the Top Italian Scientists (hfactor>30) of VIA-Academy of which 30 work in Italy.

Hence, astronomy represents an area with a typical or slightly higher than average female presence in Italian science.

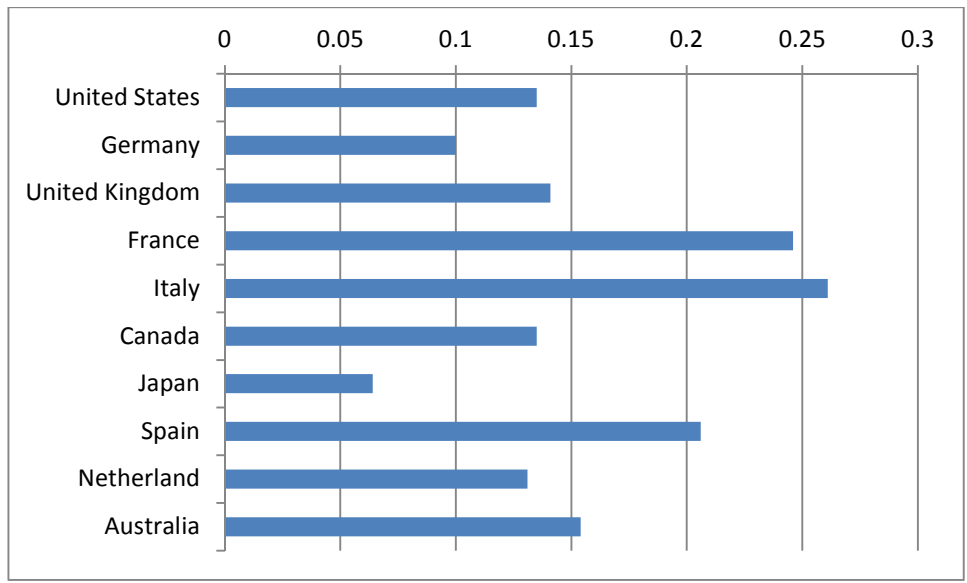

Figure 3. Fraction of women among IAU members of the ten most productive countries in astronomy

## Women in Italian astronomy

About 27% of Italian IAU members are women (146 out of 559); this is third among countries with >100 IAU members, after Argentina and Ukraine, but before France (24.6%), Spain (20.6%), China (15.3%), UK (14.1%), USA and Canada (13.5%), Netherland (13.1%), Sweden (12.6%), Germany (10.0%), Japan (6.4%)[2]. Relative to other countries, the presence of women in Italian astronomy is clearly very high: Italy has in fact the largest fraction of women astronomers among the ten world leading countries in astronomy (see Figure 3).

## Women presence in Universities

There are 47 Full Professors in Astronomy in Italian universities but only 4 of them are women (9%); 42 Associate Professors in Astronomy in Italian universities but only 6 are women (14.2%); and 72 Researchers in Astronomy in Italian Universities and 17 are women (23.6%). These fractions are similar to those for Physics for 1999 cited by Pancheri [3]: 29 out of 645 (4%) Full professors, 142 out of 963 Associate Professors (15%), and 187 out of 757 (25%) Researchers.

---

[2] Source: http://www.iau.org/administration/membership/individual/distribution/
[3] Pancheri, G., 2002, Analysis, 1

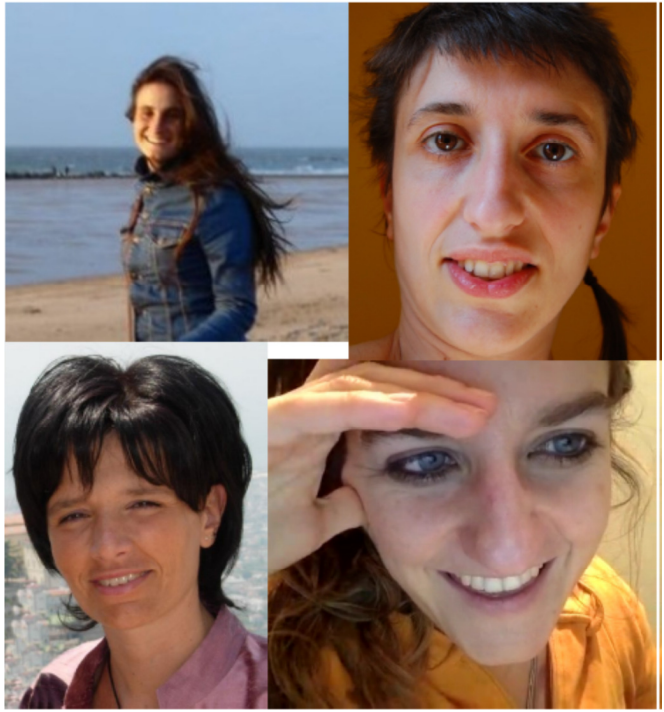

Figure 4. Women winners of the Livio Gratton Prize: Claudia Travaglio (winner in 2001), Michela Mapelli (2007), Simona Gallerani (2009), and Stefania Salvadori (2011).

## Italian women do excellent jobs: Livio Gratton Prize

The Livio Gratton Prize is awarded to the best doctoral dissertation in astronomy in Italy. Other excellent dissertations receive nominations. The Livio Gratton Prize has been assigned every two years since 1993 (11 editions). 4 out of 11 winners were women (see Figure 4); 9 out of 22 nominations were given to women. Therefore, about 40% of the best astronomy PhD theses over the last 20 years in Italy were done by women.

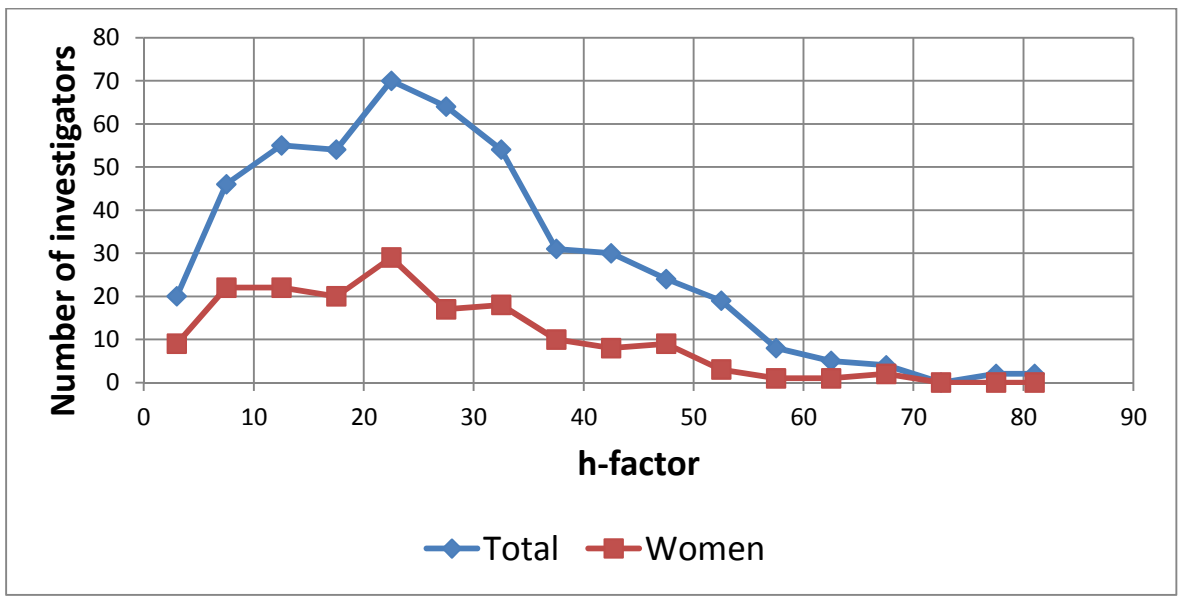

Figure 5. Distribution of h-factor for Italian astronomers, and for women astronomers in particular (Source: ADS; updated at March 2013)

## h-factor distribution for Italian astronomers

Figure 5 gives the distribution of h-factor for Italian astronomers and for women astronomers in particular[4]. This distribution is made considering all members of Macroareas 1-4 of INAF (excluding then Macroarea 5). The distribution for women follows the total one, with a peak at h=22, and a descent at higher values, somewhat steeper than for the total.

# Women researchers within INAF

## Career distribution at INAF

Women make up 32% of INAF research staff (289 out of 908) and 40% of the technical/administrative staff (173 out of 433). If we consider only permanent positions, the percentage is 30% (180 out of 599) for the research and 40% (163 out of 410) of the technical/administrative staff. For comparison, the same percentages for INFN (only permanent positions) were 22% (131 out of 603) for the research and 31% (307 out of 1004) for the technical/administrative staff at end 2011[5]. Table 1 gives the distribution among different positions at INAF[6]. Figure 6 displays photos of women that are Full Professor or Ordinario/Dirigente di Ricerca in Astronomy.

Table 1. Distribution of women among different positions at INAF (source Anagrafica INAF)

| Position | Total | Women | Fraction |
|---|---|---|---|
| **Permanent Positions** | | | |
| Ordinario/Dirigente di ricerca | 38 | 5 | 0.13 |
| Associato/Primo ricercatore | 136 | 23 | 0.17 |
| Ricercatore | 425 | 152 | 0.36 |
| **Non-permanent positions** | | | |
| Ordinario/Dirigente di ricerca | 4 | 0 | 0.00 |
| Associato/Primo ricercatore | 1 | 0 | 0.00 |
| Ricercatore | 61 | 19 | 0.31 |
| Assegno di ricerca | 176 | 64 | 0.36 |
| Borsa di studio | 67 | 26 | 0.39 |
| **Technical and Administrative staff** | | | |
| Permanent Positions | 410 | 163 | 0.40 |
| Non-permanent positions | 23 | 10 | 0.44 |

---

[4] Source: ADS (http://esoads.eso.org/abstract_service.html) updated at March 2013. We excluded members of Macroarea 5 from this statistics because people working in technology have citation rates much lower than those working in Science topics and the fraction of women is lower in this area. This might produce a bias in the results.
[5] Source: https://web2.infn.it/CUG/images/alfresco/Cug/2012/2012lStatistiche.pdf
[6] Source: http://www.ced.inaf.it/anagrafica/

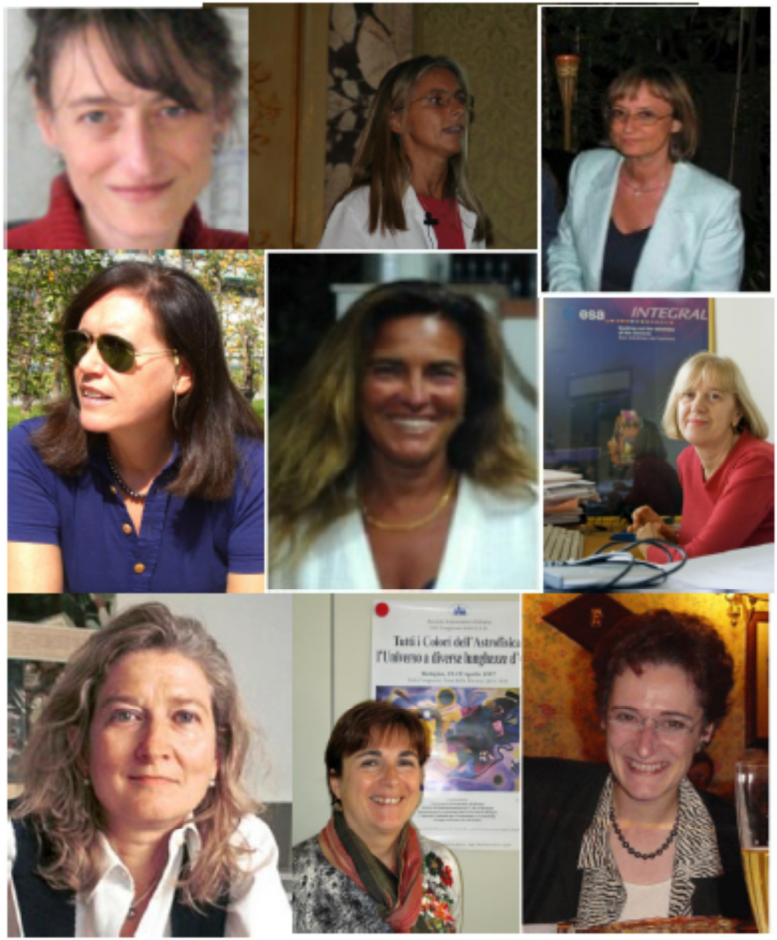

Figure 6. Women Full Professor or Ordinario/Dirigente di Ricerca in astronomy in Italy

## Distribution along Macroareas

Table 2 gives data about the women that are working in the different INAF Macroareas. We made this statistics using the mailing lists that individual Macroarea Committees have built up. The fraction of women is quite uniform at about 1/3 in the scientific Macroareas, and about half that in the technological one.

Table 2. Women presence among different INAF Macroareas

| Macroarea | Total | Women | Fraction |
| --- | --- | --- | --- |
| 1. Galaxies and Cosmology | 182 | 63 | 0.35 |
| 2. Stars and Interstellar Medium | 163 | 58 | 0.36 |
| 3. The Sun and the Solar System | 133 | 40 | 0.30 |
| 4. High Energy and Relativity | 136 | 46 | 0.34 |
| 5. Technology | 179 | 30 | 0.17 |

Table 3. Distribution of women among different positions in US astronomy Institutions[7]

| Rank | NSF | AIP | CSWA | NDS |
|---|---|---|---|---|
| Bachelor's Recipient | 43 | 40 | | |
| First-Year Grad | | 44 | | |
| Grad Student | 33 | 32 | 30 | |
| Master's Recipient | 39 | 30 | | |
| PhD Recipient | 26 | 33 | | 22 |
| Postdoc | 21 | | 22 | |
| Instructor/Adjunct | | 15 | | |
| Assistant Professor | | 28 | 20 | 20 |
| Associate Professor | | 24 | 21 | 16 |
| Full Professor | | 11 | 9 | 10 |

## Is there gender discrimination?

Career prospects for Italian astronomers are clearly worse for women than for men; only a part of this difference is due to historical reasons: higher positions are mainly occupied by older investigators and women presence was lower in the past. The low presence of women among higher career positions is similar to what found in other countries, e.g. USA, as displayed in Table 3[8]. In Germany, a mere 6.0% of the top positions at the Max Planck Society (MPS) are currently occupied by female Directors and just under 23% of posts as the head of an Independent Junior Research Group at the MPS are currently held by women scientists (MPS statistics 2007[9]). More in general (not only astronomy), in Europe, only 36% of mid-ranking professors, and 18% of full professors, are women, despite equal proportions of men and women at the undergraduate level (Vernos 2013[10]).

Is this due to an explicit discrimination (men are preferred to women at astronomy job competitions) or rather is an implicit discrimination (traditional distribution of duties within families make more difficult for women to be as productive as men during their career)?

---

[7] Source http://www.grammai.org/astrowomen/allstats.html
[8] Source http://www.grammai.org/astrowomen/allstats.html
[9] Source http://www.minerva-femmenet.mpg.de/pdf_biospektrum_legrumundhaas_engl.pdf
[10] Vernos, 2013, Nature, 495, 39

### Distribution of h-factor

We may assume that the h-factor is not as strongly affected by explicit sex discrimination as career distribution. We plotted in 7 the female fraction among researchers with an h-factor larger than a given threshold as derived from ADS[11]. This distribution appears similar to that for the career positions, with a slightly lower than average frequency of women at very high values (though the fraction is still ~20% for h>50).

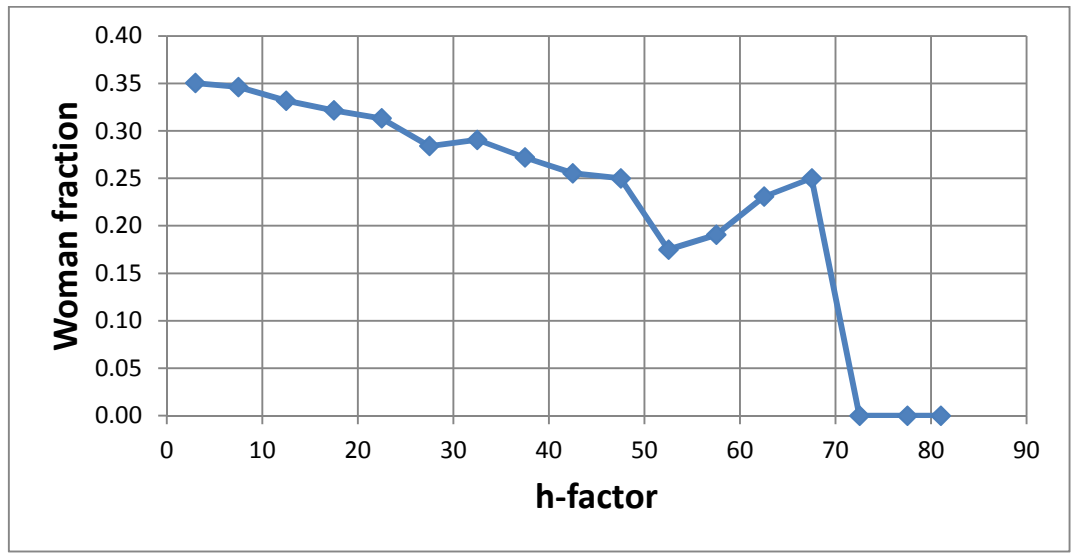

Figure 7 Women fraction among researchers with an h-factor larger than a given threshold

### Presence of women among most influential astronomers

We also notice that there are 18 women among 100 Italian most cited astronomers (ordered by normalized citations; source ADS[12]). 7 of them work outside Italy; this is to be compared with a total of Italians working outside that is 27 over 100 most cited. Women make up only 15% among 100 most cited astronomers working in Italy. This fraction is almost twice the 17 women (6 of them Italian) among the 200 most cited European astronomers (28 of them Italian).

### Presence of women among most influential papers

Since statistics on career and overall citation may be explained as a residual of a lower female presence in the past, we considered a different statistics that is only based on the last few years. Table 4 gives the fraction of Italian women first author in the first 200 most cited publications of each year over the world (still from ADS). Again, the fraction (11 out of 79) is lower than that of women among Italian astronomers, but similar to that of those in higher career positions.

## Conclusions

Italy has the highest fraction of women among the ten leading countries in astronomy. Their role is especially important in INAF, where they make up almost 30% of the research staff members. In spite of this, women are

---

[11] Source: ADS (http://esoads.eso.org/abstract_service.html ) updated at March 2013. We excluded members of Macroarea 5 from this statistics because people working in technology have citation rates much lower than those working in Science topics and the fraction of women is lower in this area. This might produce a bias in the results.

[12] Source: http://esoads.eso.org/abstract_service.html

under-represented in higher career positions. Trying to understand these facts, we observe that the distribution of h-factors and the statistics about most cited papers show that:

Table 4. Fraction of Italian women first author in the first 200 most quoted publications of each year over the world

| Year | Total | No. women |
|---|---|---|
| 2006 | 12 | 1 |
| 2007 | 8 | 2 |
| 2008 | 17 | 2 |
| 2009 | 13 | 2 |
| 2010 | 13 | 3 |
| 2011 | 8 | 1 |
| 2012 | 8 | 0 |
| Total | 79 | 11 |

- Female presence is also lower among researchers with high h-factors and among first-authors of most influential papers. The run is quite similar to that observed for the career presence

- This suggests that implicit sex discrimination factors likely dominate over explicit ones and are still at work (see also Vernos 2013, for a similar result concerning ERC grants). Note however that explicit gender discrimination factors are considered to be very important by many – see e.g. the responses to the questionnaire on European Platform of Women Scientists [13]. INAF explicitly acts to correct such discrimination by taking care to have adequate female presence in Committees (2 out of 5 in the Council; 3 out of 7 in the Scientific Committee; 4 out of 17 Directors of INAF Institutes) and Panels, as monitored by the Comitato Unico di Garanzia. INAF vice-president and the Chair of the Scientific Committee are women.

Moreover, we wonder why the countries with the highest fraction of women among astronomers are the catholic countries (Argentina, Italy, France, Spain, Portugal, Brazil etc.) and not the protestant ones where traditionally the women position is considered to be better (Sweden, Denmark, Netherland, Germany, UK, USA: see Figure 8), and what are the implicit sex discrimination factors that are still so strong, given that the explicit ones are likely weakening with time.

For us and many others[14], the discriminator factor is obvious, that is, the much larger burden due to child care that is not equally distributed among the sexes in a typical western country (see also Shen 2013[15]): this is only partially compensated by job laws over the world, even in most advanced countries. This makes it difficult for a woman to make a scientific career, unless she either has an unusually strong support from her family or chooses not to have children or has only one.

---

[13] http://home.epws.org/2289331/The-Excellence-debate-a-round-up-of-responses-thus-far
[14] Three decades ago, when Nobel laureate Rosalyn Yalow spoke to a "women in science" group at a major university, her opening statement was: "The primary problem is childcare. Everything else is secondary." (see Barres in Nature, 495, 35, 2013)
[15] Shen, 2013, Nature 495, 22

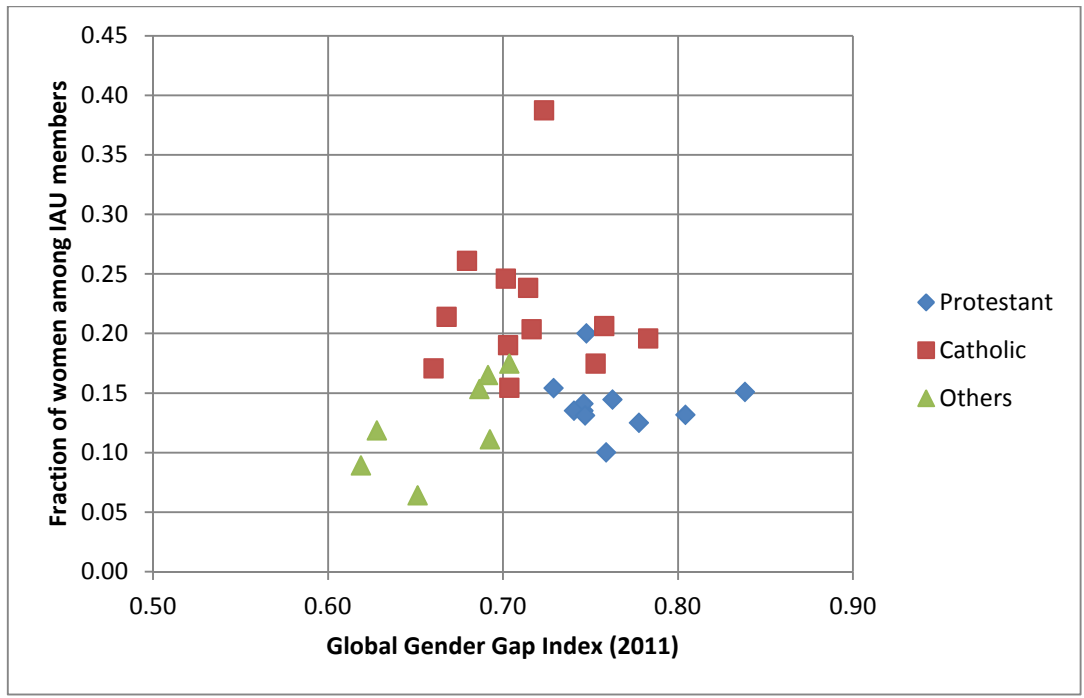

Figure 8. Relation between the fraction of women among IAU members and the Global Gender Gap Index for Year 2011[16] for the thirty countries contributing more to astronomy worldwide. Note that a higher value of the Gender Gap Index indicates less discrimination against women. Different symbols are used for countries of different culture/religion.

However, there should be something special in the Latin-countries that determine the high fraction of women astronomers. This higher presence of women is not limited to astronomy (see the data cited by Pancheri 2002 about Italy and Portugal, and the comparison with analogous data for UK). We suggest that this probably depended on the peculiar young age (25 to 35) at which in the Latin countries astronomers could get a permanent position in the recent past (up to some ten years ago; this is not any more the case, at least in Italy). This implied that many more women could combine having children with a job in science, often however at the expense of accepting a subordinate role rather than being a group leader. In contrast, the long periods with non-permanent position which are a tradition in Anglo-Saxon and northern Europe science (and now holding also in Italy) imply that only women with extraordinary qualities or very career-oriented may become scientists. We then leave to the reader to draw her/his conclusions about the merits of the current career pattern in Italy, where permanent positions are now acquired at an age >35 (more typically >40) after a long period of very short-term contracts. We however fear that no significant improvement about the role of women in astronomy can be achieved unless a very different career pattern is adopted. We finally recommend that there should not be any educational discrimination since pre-scholar ages and that girls should be encouraged to follow their will if they wish to follow careers traditionally considered more "masculine".

---

[16] Source http://www3.weforum.org/docs/WEF_GenderGap_Report_2011.pdf